\documentclass[prb,twocolumn,amsmath,amssymb,superscriptaddress,showpacs]{revtex4-1}

\usepackage{bm}
\usepackage{graphicx}
\usepackage[bookmarks=false]{hyperref}

\usepackage{latexsym}
\usepackage{color}

\newcommand{\bra}[1]{\langle #1|}
\newcommand{\ket}[1]{|#1\rangle}
\newcommand{\braket}[1]{\langle #1 \rangle}

\def\dd{\mathrm{d}}
\def\ee{\mathrm{e}}
\def\ii{\mathrm{i}}

\def\Hc{\mathrm{H.c.}}

\def\ddt#1{\frac{\dd #1}{\dd t}}

\def\Tr{\mathrm{Tr}}

\def\eV{\mathrm{eV}}
\def\meV{\mathrm{meV}}

\def\nm{\mathrm{nm}}
\def\micron{\mu\mathrm{m}}
\def\sec{\mathrm{s}}
\def\ps{\mathrm{ps}}
\def\AA{\mathrm{\mathring{A}}}

\def\wx{\varOmega}
\def\rabiP{g}
\def\dm{d}
\def\sct{\mathcal{V}}
\def\vol{V}
\def\volunit{V_0}
\def\volcoh{V_{\text{coh}}}
\def\lencoh{\lambda_{\text{coh}}}
\def\lensct{\lambda_{\text{screen}}}
\def\diez{\varepsilon_0}
\def\diebg{\varepsilon_{\text{bg}}}
\def\die{\varepsilon}
\def\cbg{v}
\def\vg{v_{\text{g}}}
\def\thick{L}
\def\thickeff{L_{\text{eff}}}
\def\kp{k_{\parallel}}

\def\tRabi{\tau_{\text{Rabi}}}
\def\tescape{\tau_{\text{escape}}}
\def\temit{\tau_{\text{emit}}}
\def\tconv{\tau_{\text{conv}}}

\def\gammadeph{\gamma^{\text{deph}}}
\def\gammadepha{\gamma_{\text{deph}}}
\def\gammabulk{\gamma_{\text{escape}}}

\def\oH{\hat{H}}

\def\ovP{\hat{\bm{P}}}

\def\ovDT{\hat{\bm{D}}_{\perp}}
\def\oa{\hat{a}}
\def\oad{\hat{a}^{\dagger}}
\def\op{\hat{p}}
\def\opd{\hat{p}^{\dagger}}
\def\osigma{\hat{\sigma}}
\def\osigmad{\hat{\sigma}^{\dagger}}
\def\oalpha{\hat{\alpha}}

\def\obeta{\hat{\beta}}

\def\orho{\hat{\rho}}

\def\vunit{\bm{e}}
\def\vr{\bm{r}}
\def\vR{\bm{R}}
\def\vk{\bm{k}}
\def\vq{\bm{q}}

\begin{document}


\title{Theory of lifetime of exciton incoherently created below its resonance frequency
by inelastic scattering}

\author{Motoaki Bamba}
\altaffiliation{Present address:
Department of Materials Engineering Science, Osaka University,
1-3 Machikaneyama, Toyonaka, Osaka 560-8531, Japan\\
E-mail: bamba@qi.mp.es.osaka-u.ac.jp}
\affiliation{Department of Physics, Graduate School of Science, Osaka University, 1-1 Machikaneyama, Toyonaka, Osaka 560-0043, Japan}
\author{Shuji Wakaiki}
\affiliation{Department of Material and Life Science, Division of Advanced Science and Biotechnology, Graduate School of Engineering, Osaka University, 2-1 Yamada-oka, Suita, Osaka 565-0871, Japan}
\author{Hideki Ichida}
\affiliation{Department of Material and Life Science, Division of Advanced Science and Biotechnology, Graduate School of Engineering, Osaka University, 2-1 Yamada-oka, Suita, Osaka 565-0871, Japan}
\affiliation{Science and Technology Entrepreneurship Laboratory, Osaka University, 2-1 Yamada-oka, Suita, Osaka 565-0871, Japan}
\author{Kohji Mizoguchi}
\affiliation{Department of Physical Science, Graduate School of Science, Osaka Prefecture University, 1-1 Gakuen, Naka-ku, Sakai, Osaka 599-8531, Japan}
\author{DaeGwi Kim}
\affiliation{Department of Applied Physics, Graduate School of Engineering, Osaka City University, 3-3-138 Sugimoto, Sumiyoshi-ku, Osaka 558-8585, Japan}
\author{Masaaki Nakayama}
\affiliation{Department of Applied Physics, Graduate School of Engineering, Osaka City University, 3-3-138 Sugimoto, Sumiyoshi-ku, Osaka 558-8585, Japan}
\author{Yasuo Kanematsu}
\affiliation{Department of Material and Life Science, Division of Advanced Science and Biotechnology, Graduate School of Engineering, Osaka University, 2-1 Yamada-oka, Suita, Osaka 565-0871, Japan}
\affiliation{Science and Technology Entrepreneurship Laboratory, Osaka University, 2-1 Yamada-oka, Suita, Osaka 565-0871, Japan}

\date{\today}

\begin{abstract}
When an exciton in semiconductor is scattered
and its energy is decreased far below the resonance energy of the bare exciton state,
it has been considered that an exciton-polariton
is created immediately by the scattering process,
because there is no exciton level at that energy.
However, according to the recent time-resolved measurements of P emission
originating from inelastic exciton-exciton scattering,
it looks rather natural to consider that
the exciton-polariton is created in a finite time scale
which is restricted by a coherence volume of the exciton after the scattering.
In this interpretation, the exciton remains in this time scale
far below its resonance energy
as a transient state in a series of processes.
We propose an expression of the P-emission lifetime
depending on the coherence volume of the scattered excitons
through the conversion process from them to the polaritons.
The coherence volume of the scattered excitons appears
in the calculation of the inelastic scattering process
on the assumption of a finite coherence volume of the bottleneck excitons.
Time-resolved optical-gain measurements
could be a way for investigating the validity of our interpretation.
\end{abstract}




\maketitle
\section{Introduction}
We can obtain a variety of properties of condensed matters
from luminescence spectra
by varying sample temperature, pumping frequency, pumping intensity, etc.
\cite{klingshirn05}
Time-resolved luminescence measurements
give us more detailed information especially
about relaxation processes of the excitations
such as excitons and polaritons.
However, theoretical studies of the luminescence (spontaneous emission of the excitations)
is not yet well developed probably due to the complexity of the relaxation dynamics
involving spatial inhomogeneities, impurities, phonon scattering, spatial diffusion,
inter-excitation scattering, and so on.
The relaxation, dissipation, and dephasing processes
have been investigated mainly by nonlinear optical responses
such as pump-probe and four-wave mixing experiments.
However, even by such measurements, 
it is still hard to obtain the complete knowledge of the luminescence process,
especially the coherence volume of the excitation,
which governs the emission lifetime.
\cite{Feldmann1987PRL,Grad1988PRA,nakamura89,itoh90}

Concerning the spontaneous emission of excitations at quasi-equilibrium
(equilibrium only in matters excluding the radiation field),
the relation between the emission lifetime and the homogeneous spectral linewidth
(reflecting the coherence volume)
has been investigated for quasi-two-dimensional excitons in GaAs/AlGaAs quantum wells.
\cite{Feldmann1987PRL}
The coherence volume also gives the limit of
the so-called exciton superradiance
(size-enhancement of radiative decay rate
or of oscillator strength),\cite{Grad1988PRA,hanamura88,nakamura89,itoh90}
by which the emission lifetime is shortened with an increase in
interaction volume between the radiation field
and the center-of-mass wavefunction of excitons (radius of quantum dot).
There were also attempts for estimating theoretically
the coherence volume of excitations
such as by dephasing rate.\cite{Grad1988PRA}
However, the understanding of the coherence volume
is not yet well developed,
because it is usually estimated only through the emission lifetime
and the luminescence is in fact influenced by many other processes and factors,
such as reabsorption of photons,
stimulated emission of photons,
diffusion of excitation,
ballistic propagation of photons,
penetration depth of pumping (spatial inhomogeneity),
internal reflection, etc.\cite{klingshirn05}

\begin{figure}[tbp]
\begin{center}
\includegraphics[width=.6\linewidth]{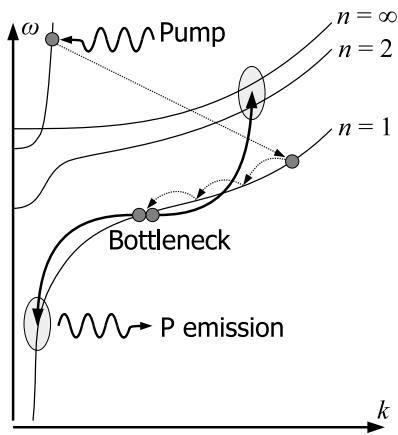}
\caption{Sketch of P-emission process depicted in dispersion relations.
Excitons created by pumping are relaxed to the bottleneck region
(dashed arrows).
Then, they are scattered to higher exciton states
with $n > 1$ and to photon-like polariton states with conserving the energy.
The emission with the lower energy is called the P emission.}
\label{fig:1}
\end{center}
\end{figure}

Although the emission frequency is almost fixed
for the spontaneous emission of excitons at the quasi-equilibrium
(called the bottleneck region
\cite{Toyozawa1959PTPS,Heim1973PRL,Sumi1976JPSJ}
in the picture of exciton-polaritons),
we can also observe luminescence peaks at lower frequencies,
which involve the emission of optical phonons,
inelastic exciton-exciton scattering (P emission),
exciton-carrier scattering (H emission),
and excitonic molecules (M emission).
\cite{Klingshirn1981PR,klingshirn05}
In the P-emission process,
one exciton is inelastically scattered to a higher exciton state
and the other one is scattered to the photon-like polariton branch
as depicted in Fig.~\ref{fig:1}.
It emerges under high-power pumping exceeding a threshold,
and we have also an optical gain at the P-emission frequency.
\cite{Cingolani1982PRB,Tang1998APL,Ichida2000JL,Tanaka2002JAP,Nakayama2005APL,Hashimoto2010APL}
The relaxation and scattering processes toward the P emission
has been investigated in time-resolved measurements performed
by optical Kerr gating method
\cite{Ichida2005PRB,Takeda2006JJAP,Ichida2007PRB,Hirano2010PRB,Wakaiki2011pssc,Ando2012PRB,Wakaiki2013EPJB,Wakaiki2014JL,Wakaiki2015PE}
and by streak camera,\cite{Nakayama2008APL,Furukawa2014JPSJ}
and then the following facts have been revealed.
1) The onset time reflects the time of energy relaxation of excitons
toward the bottleneck region on the lower exciton-polariton branch.
\cite{Takeda2006JJAP,Ichida2007PRB,Ando2012PRB}
2) The rise time reflects the rate of the inelastic scattering of excitons.
\cite{Ichida2005PRB}
3) The peak frequency is changed temporally during the rise and decay periods.
\cite{Ichida2005PRB,Nakayama2008APL,Ando2012PRB}
Whereas this fact could be interpreted as
the change of effective temperature
(distribution) of excitons at the bottleneck,
we can also interpret it as
that the decay time (lifetime) of the P emission at each emission frequency
strongly depends on that frequency.

The typical P-emission lifetimes are observed as
a few ps\cite{Ichida2005PRB,Takeda2006JJAP,Ichida2007PRB,Hirano2010PRB,Wakaiki2011pssc,Wakaiki2013EPJB,Wakaiki2014JL,Wakaiki2015PE}
or a few tens of ps.\cite{Nakayama2008APL,Ando2012PRB,Furukawa2014JPSJ}
Although these lifetimes basically depend on materials of samples,
they are generally much shorter than the emission lifetime $\temit$
of bottleneck excitons at quasi-equilibrium
(in the order of nanoseconds).
The time-resolved measurements revealed also that
the P-emission lifetime is an increasing function of the emission frequency,
\cite{Hirano2010PRB,Wakaiki2011pssc,Wakaiki2013EPJB,Wakaiki2014JL,Wakaiki2015PE,Furukawa2014JPSJ}
and the lifetime at each emission frequency
is almost independent of the pumping power.\cite{Wakaiki2014JL,Wakaiki2015PE}
Note that the emission-frequency-dependence of the P-emission lifetime
can be scaled phenomenologically by that of inverse of the group velocity
of the photon-like polariton.\cite{Wakaiki2011pssc,Wakaiki2013EPJB,Wakaiki2014JL,Wakaiki2015PE}
The lifetime of the spectrally-integrated P-emission signal was shortened
through the lowering of the peak frequency
with an increase in pumping power (effective temperature)
for InGaN.\cite{Nakayama2008APL}
However, it was almost unchanged for CuI\cite{Ichida2005PRB,Ichida2007PRB}
and AlGaN,\cite{Hirano2010PRB}
because the peak frequencies were not strongly changed.
Further, the lifetime of the spectrally-integrated P-emission signal
was also independent of the pumping frequency for CuI.\cite{Ichida2007PRB}

The P emission at each emission frequency shows an exponential decay in time.
Its decay time is independent of the pumping power,
and depends strongly on the emission frequency
(inversely proportional to group velocity).
Then, it is now recognized that
the P-emission lifetime does not reflect the lifetime of excitons at the bottleneck,
but it rather reflects the lifetime of quasi-particles
(excitons or exciton-polaritons) after the inelastic scattering.
In Ref.~\onlinecite{Ichida2005PRB}, the authors concluded that
it reflects the lifetime of photon-like polaritons,
which is considered to be shortened by the increase of photonic fraction
of the polariton state.
However, from the sample thickness and the group velocity of polaritons,
the lifetime (escape time) of photon-like polaritons is
estimated to be much shorter (tens of femtoseconds)
than the P-emission lifetimes (picoseconds) observed in experiments.
\cite{Wakaiki2011pssc,Wakaiki2013EPJB}
On the other hand, in Ref.~\onlinecite{Hirano2010PRB},
the authors analyzed the P-emission decay
as diffusive propagation of the photon-like polaritons,
although the diffusion of light is usually discussed
in strongly disordered media,
where excitons should lose the memory of propagation direction quickly
compared to the reemission time scale.

In this paper, from the viewpoint of the coherence volume,
we try to propose the following interpretation
of the P-emission lifetime:
Just after the inelastic exciton-exciton scattering,
the photon-like polariton is not immediately created,
but the exciton remains with losing its energy
in a time scale of picoseconds as depicted in Fig.~\ref{fig:2}.
Then, the conversion time from the exciton to photon-like polariton,
which is restricted by the coherence volume,
is observed as the P-emission lifetime.
Although the P emission has been considered
as a stimulated emission of polaritons,
\cite{klingshirn05,Cingolani1982PRB,Tang1998APL,Ichida2000JL,Tanaka2002JAP,Nakayama2005APL,Hashimoto2010APL}
we need to reconsider it as a stimulated creation (scattering) of excitons
in our interpretation.

In Sec.~\ref{sec:expression},
we first estimate the
interchange time between exciton and photon in the polariton states,
the radiative recombination time of excitons,
and the escape time of the polaritons.
Only the radiative recombination time depends on the coherence volume.
In Sec.~\ref{sec:P-emission},
we explain the detail of our interpretation of the P emission
after the inelastic exciton-exciton scattering.
Its justification and further discussion are performed
in Sec.~\ref{sec:discussion}.
The summary is shown in Sec.~\ref{sec:summary}.

\section{Characteristic time scales of excitons and polaritons} \label{sec:expression}
We first calculate the exciton-photon interchange time in polariton states
and the radiative recombination time of exciton
from the Hamiltonian of light-matter coupling.
We consider an homogeneous background medium with a relative dielectric constant $\diebg$,
and the Hamiltonian of the radiation field in the background medium is written as
\begin{equation}
\oH_{\text{rad}}
= \sum_{\eta=1,2} \sum_{\vk} \hbar \cbg|\vk| \oad_{\vk,\eta} \oa_{\vk,\eta},
\end{equation}
where $\oa_{\vk,\eta}$ is the annihilation operator of a photon with wavevector $\vk$
and polarization $\eta$,
and $\cbg = c / \sqrt{\diebg}$ is the speed of light in the background medium
for the speed $c$ in vacuum.
The Hamiltonian of the light-matter coupling is expressed in the electric-dipole gauge as
\cite{cohen-tannoudji89,Todorov2014PRB,Bamba2014SPT}
\begin{equation} \label{eq:oHLM} 
\oH_{\text{LM}}
= - \frac{1}{\diez\diebg}\int\dd\vr\ \ovP(\vr) \cdot \ovDT(\vr).
\end{equation}
Here, the transverse component of the electric displacement field is defined
\begin{equation}
\ovDT(\vr) = \sum_{\eta=1,2} \sum_{\vk} \vunit_{\vk,\eta}
  \ii\sqrt{\frac{\hbar\diez\diebg\cbg|\vk|}{2\vol}}
  \left( \oa_{\vk,\eta} - \oad_{-\vk,\eta} \right) \ee^{\ii\vk\cdot\vr},
\end{equation}
where $\vunit_{\vk,\eta}$ is the unit vector perpendicular to $\vk$,
and $\vol$ is the volume of the space.
$\ovP(\vr)$ represents the polarization density involving the creation and annihilation of
an electron-hole pair as
\begin{equation}
\ovP(\vr) = \dm_{cv} \sum_{\xi} \vunit_{\xi}
\sum_{\lambda} \delta(\vr-\vR_{\lambda})
\left(\oalpha_{\xi,\vR_\lambda}\obeta_{\xi,\vR_\lambda} + \Hc\right).
\end{equation}
Here, $\dm_{cv}$ is the transition dipole moment
calculated under the long-wavelength approximation,
$\vunit_{\xi}$ is the unit vector in the direction $\xi = \{x, y, z\}$,
and $\vR_{\lambda}$ is the position of unit cell $\lambda$.
$\oalpha_{\xi,\vR_\lambda}$ and $\obeta_{\xi,\vR_\lambda}$ are, respectively,
annihilation operators of an electron and a hole at $\vR_\lambda$
involving the polarization in the $\xi$ direction.
The optical inter-band transition is supposed to occur
almost inside a unit cell.
The annihilation operator of an exciton in state $\mu$
with a wavefunction $\psi_{\mu}(\vr)$ of the electron-hole relative motion
and a center-of-mass at $\vR_{\lambda}$ is written as
\begin{equation}
\osigma_{\mu,\lambda} = \int\dd\vr\ \frac{\psi_{\mu}(\vr)}{\sqrt{\volunit}}
\oalpha_{\xi_{\mu},\vR_\lambda+(m_h/M)\vr}\obeta_{\xi_{\mu},\vR_\lambda-(m_e/M)\vr}.
\end{equation}
Here, $m_e$ and $m_h$ are the effective mass of the electron and hole,
respectively, and $M = m_e + m_h$ is the total mass.
$\volunit$ is the volume of a unit cell.
The wavefunction is normalized as
$\int\dd\vr\ \psi_{\mu}(\vr)^*\psi_{\mu'}(\vr) = \delta_{\mu,\mu'}$.
Due to the completeness of the exciton state
$\sum_{\mu}\psi_{\mu}(\vr)^*\psi_{\mu}(\vr') = \delta(\vr-\vr')$,
the polarization density is rewritten as\cite{haug04}
\begin{equation}
\ovP(\vr) = \dm_{cv} \sum_{\mu} \vunit_{\mu} \sqrt{\volunit} \psi_{\mu}(0)
\sum_{\lambda} \delta(\vr-\vR_{\lambda})
\left( \osigma_{\mu,\lambda} + \osigmad_{\mu,\lambda} \right).
\end{equation}
Here, we defined the wavefunction $\psi_{\mu}(\vr)$
to be real at $\vr = 0$.
The degree of freedom of the polarization direction $\xi$
is included to the index $\mu$ of the exciton state.
Using this expression, Eq.~\eqref{eq:oHLM} is rewritten as
\begin{align} \label{eq:oHLM2} 
\oH_{\text{LM}}
& = - \sum_{\mu} \sum_{\eta=1,2} \sum_{\vk} \vunit_{\mu} \cdot \vunit_{\vk,\eta}
  \ii\sqrt{\frac{\hbar\cbg|\vk|\dm_{cv}{}^2\psi_{\mu}(0)^2}{2\diez\diebg N}}
\nonumber \\ & \quad \times
  \sum_{\lambda}
  \left( \osigma_{\mu,\lambda} + \osigmad_{\mu,\lambda} \right)
  \left( \oa_{\vk,\eta} - \oad_{-\vk,\eta} \right) \ee^{\ii\vk\cdot\vR_{\lambda}}.
\end{align}
Here, $N = \vol/\volunit$ is the number of the unit cells in the whole space.
For Bohr radius $a_B^*$ of the exciton larger enough than the lattice constant
(in the limit of Wannier exciton),\cite{haug04}
the $s$-orbital
wavefunction
of the electron-hole relative motion is expressed as
\begin{equation}
\psi_{ns}(0) = \sqrt{\frac{1}{\pi a_B^*{}^3} \frac{1}{n^3}}.
\end{equation}

\subsection{Exciton-photon interchange time} \label{eq:interchange}
We define the exciton operator in the $\vk$-representation as
\begin{equation}
\osigma_{\mu,\vk}
= \frac{1}{\sqrt{N}} \sum_{\lambda} \ee^{-\ii\vk\cdot\vR_{\lambda}} \osigma_{\mu,\lambda}.
\end{equation}
The Hamiltonian of the excitons is represented as
\begin{equation}
\oH_{\text{ex}}
= \sum_{\mu}\sum_{\vk}\hbar\wx_{\mu,k}\osigmad_{\mu,\vk}\osigma_{\mu,\vk}
  + \frac{1}{2\diez\diebg}\int\dd\vr\ \ovP(\vr)\cdot\ovP(\vr).
\end{equation}
Here, $\wx_{\mu,k}$ is the eigenfrequency of exciton in state $\mu$ with wavenumber $k$.
The last term is the so-called $P^2$ term
and represents the depolarization shift.\cite{cohen-tannoudji89,Todorov2014PRB,Bamba2014SPT}
The light-matter coupling Hamiltonian given by Eq.~\eqref{eq:oHLM2} is rewritten as
\begin{align} \label{eq:oHLM3} 
\oH_{\text{LM}}
& = - \sum_{\mu} \sum_{\eta=1,2} \sum_{\vk}
  \vunit_{\mu} \cdot \vunit_{\vk,\eta}
\nonumber \\ & \quad \times
  \ii\hbar\rabiP_{\mu,k}
  \left( \osigma_{\mu,-\vk} + \osigmad_{\mu,\vk} \right)
  \left( \oa_{\vk,\eta} - \oad_{-\vk,\eta} \right),
\end{align}
where the coupling strength is defined as
\begin{equation} \label{eq:rabiP} 
\rabiP_{\mu,k} = 
  \sqrt{\frac{\cbg k\dm_{cv}{}^2\psi_{\mu}(0)^2}{2\hbar\diez\diebg}}.
\end{equation}
When the Fermionic nature of the exciton can be neglected
in the one-body problem with respect to the exciton,
the eigenstates of the electromagnetic fields in this excitonic medium
are the polariton states satisfying the dispersion relation
(roughly sketched in Fig.~\ref{fig:1}) as
\begin{equation} \label{eq:dispersion} 
\frac{c^2k^2}{\omega^2}
= \diebg + \sum_{\mu} \frac{4\pi\beta_{\mu,k}\wx_{\mu,k}{}^2}{\wx_{\mu,k}{}^2-(\omega+\ii0^+)^2}
= \die(\omega,k),
\end{equation}
where the non-dimensional factor is defined as
\begin{equation}
4\pi\beta_{\mu,k} = \frac{4\diebg\rabiP_{\mu,k}{}^2}{\wx_{\mu,k}\cbg k}
= \frac{2\dm_{cv}{}^2\psi_{\mu}(0)^2}{\diez\hbar\wx_{\mu,k}}.
\end{equation}
When the polaritons exist stably in a large enough medium
with negligible dissipation,
the interchange rate between exciton state $\mu$ and photon one is estimated
from Eq.~\eqref{eq:rabiP} for $k = \wx_{\mu} / \cbg$ as
\begin{equation} \label{eq:rabiP_mu} 
\rabiP_{\mu} = \sqrt{\frac{\wx_{\mu,k}\dm_{cv}{}^2\psi_{\mu}(0)^2}{2\hbar\diez\diebg}}
= \sqrt{\frac{\pi\beta_{\mu}\wx_{\mu}{}^2}{\diebg}}.
\end{equation}
For A exciton state with $n = 1$ ($1s$) in ZnO,\cite{Madelung2001}
we have $\wx_{A,1s} = 3.375\;\eV$, $\diebg = 4$,
and $\varDelta_{A,1s} = 4\pi\beta_{A,1s}\wx_{A,1s}/\diebg = 5.74\;\meV$
($\wx_{B,1s} = 3.381\;\eV$ and $\varDelta_{B,1s} = 6.62\;\meV$ for B exciton).
The interchange rate is then estimated as $\hbar\rabiP_{A,1s} =70\;\meV$
($\hbar\rabiP_{B,1s} = 75\;\meV$).
The interchange time $\tRabi = 2\pi/\rabiP_{\mu} = 0.06\;\ps$
is one or two orders of magnitude shorter than the P-emission lifetime
observed in the experiments.
\cite{Wakaiki2011pssc,Wakaiki2013EPJB,Wakaiki2014JL}

\subsection{Radiative recombination time of exciton} \label{sec:radiative_decay}
Let us next calculate the radiative recombination rate of excitons
from the light-matter-coupling Hamiltonian \eqref{eq:oHLM2}.
Here, we suppose an exciton in state $\mu$ as an initial state
and its center-of-mass is localized at $\vR_{\lambda}$.
According to the Fermi's golden rule,
the transition rate from the exciton state
to one photon state with any $\vk$ and $\eta$ is obtained as
\begin{align}
\gamma_{\mu}
& = \frac{2\pi}{\hbar}\sum_{\eta,\vk}
    |\braket{0|\oa_{\vk,\eta}\oH_{\text{LM}}\osigmad_{\mu,\lambda}|0}|^2
    \delta(\hbar\wx_{\mu}-\hbar\cbg|\vk|) \nonumber \\
&
= \frac{\wx_{\mu}{}^3\dm_{cv}{}^2\psi_{\mu}(0)^2\volunit}{3\pi\hbar\diez\diebg\cbg{}^3}
  = \frac{2\rabiP_{\mu}{}^2\wx_{\mu}{}^2\volunit}{3\pi\cbg{}^3},
\label{eq:gamma} 
\end{align}
where $\ket{0}$ is the vacuum state
and we used the following relation for arbitrary function $F(k)$
\begin{equation}
\sum_{\eta=1,2}\int\dd\vk\ | \vunit_{\mu}\cdot\vunit_{\vk,\eta} |^2 F(|\vk|)
= \int_0^{\infty}\dd k\ \frac{8\pi k^2}{3}  F(k).
\end{equation}
For ZnO, the lattice constants are $a = 3.25\AA$ and $c = 5.21\AA$,
\cite{Madelung2001}
and then the volume of the unit cell is
\begin{equation}
\volunit = \frac{1}{2} \frac{\sqrt{3}}{2}\times(3.25\AA)^2\times5.21\AA
= 24 \AA^3.
\end{equation}
Therefore, the radiative recombination rate \eqref{eq:gamma} is estimated for the A excitons as
\begin{equation} \label{eq:gamma_A1s} 
\gamma_{A,1s} = 0.45\;(\mu\sec)^{-1}.
\end{equation}
This rate is quite low even compared to
the spontaneous emission rate $1/\temit$ of bottleneck excitons
observed in luminescence experiments (usually in the order of nanoseconds).

This is because the center-of-mass of exciton is in fact not localized
at a unit cell, but it coherently spreads in a finite volume,
which is called the coherence volume $\volcoh^{\mu}$.
Here, we suppose that such exciton state in state $\mu$ is represented
with a center-of-mass wavefunction $\varPhi_{\mu}(\vr)$ as
\begin{equation} \label{eq|ex_mu>} 
\ket{\text{ex}_{\mu}} = \sum_{\lambda} \sqrt{\volunit} \varPhi_{\mu}(\vR_{\lambda}) \osigmad_{\mu,\lambda}\ket{0},
\end{equation}
where the wavefunction is normalized as $\int\dd\vr\ |\varPhi_{\mu}(\vr)|^2 = 1$.
Instead of Eq.~\eqref{eq:gamma},
the radiative recombination rate of this exciton state is written as
\begin{align}
\varGamma_{\mu}
& = \frac{2\pi}{\hbar}\sum_{\eta,\vk}
    |\braket{0|\oa_{\vk,\eta}\oH_{\text{LM}}|\text{ex}_{\mu}}|^2
    \delta(\hbar\wx_{\mu}-\hbar\cbg|\vk|), \nonumber \\
& = \frac{2\pi}{\hbar}\sum_{\eta,\vk}
    \left|
      \braket{0|\oa_{\vk,\eta}\oH_{\text{LM}}\osigmad_{\mu,\lambda}|0}
      \phi_{\mu,\vk}
    \right|^2
    \delta(\hbar\wx_{\mu}-\hbar\cbg|\vk|),
\label{eq:Gammma_mu} 
\end{align}
where the factor $\phi_{\mu,\vk}$ is defined as
\begin{equation}
\phi_{\mu,\vk}
= \sum_{\lambda} \ee^{\ii\vk\cdot\vR_{\lambda}}
  \sqrt{\volunit} \varPhi_{\mu}(\vR_{\lambda})
= \int\dd\vr\ \frac{\ee^{\ii\vk\cdot\vr}}{\sqrt{\volunit}} \varPhi_{\mu}(\vr).
\end{equation}
Here, we suppose that the coherence length $(\volcoh^{\mu})^{1/3}$
is shorter enough than the radiation wavelength $2\pi/|\vk|$.
Further, the amplitude of the center-of-mass wavefunction $\varPhi_{\mu}(\vr)$
is supposed to be almost homogeneous,
i.e., $\varPhi_{\mu}(\vr) = 1/\sqrt{\volcoh^{\mu}}$
in the coherence volume $\volcoh^{\mu}$.
Then, $\phi_{\vk}$ does not depend on $\vk$,
and its absolute value is estimated as
\begin{equation}
|\phi_{\mu,\vk}| = \int\dd\vr\ \frac{\varPhi_{\mu}(\vr)}{\sqrt{\volunit}}
= \sqrt{\frac{\volcoh^{\mu}}{\volunit}}.
\end{equation}
Substituting this into Eq.~\eqref{eq:Gammma_mu},
the radiative recombination rate of excitons in state $\mu$
with the coherence volume $\volcoh^{\mu}$ is obtained as
\begin{equation}
\varGamma_{\mu} = \gamma_{\mu} \frac{\volcoh^{\mu}}{\volunit}.
\end{equation}
In this way, we get the size-enhancement of the radiative recombination rate
(in other words, that of oscillator strength),
and it is called the exciton superradiance.
\cite{nakamura89,itoh90,hanamura88,Grad1988PRA}
When the coherence length $(\volcoh^{\mu})^{1/3}$ is comparable to or larger
than the wavelength of the radiation,
we have to consider the crossover to the polariton picture.\cite{bamba09crossover}

Note that the interchange time $\tRabi$
[also the dispersion relation \eqref{eq:dispersion}] is obtained
without the concept of the coherence volume.
This means that all the atoms associate with each other coherently for the interchange,
while only the atoms in the coherence volume associate for the emission from localized exciton.
In other words, the interchange reflects the coherence volume of the electromagnetic fields
(widely spread by the propagation),
while the spontaneous emission reflects that of bare exciton.
Once a photon is emitted from the bare exciton,
it then gets a spatial coherence by propagating in the medium as a polariton,
if dissipations and dephasing are weak enough compared to the light-matter coupling.
This idea is important to understand the lifetime of the P emission
in the next section.

\subsection{Escape time of polariton}
We next consider another time scale, the escape time of polaritons.
We suppose a film of the excitonic medium with a thickness $\thick$,
and it is thick enough compared to the radiation wavelength.
When polariton states are supposed to be a good quantum state,
the escape time of polariton can be estimated from its group velocity
$\vg = \partial\omega/\partial k$,\cite{shuh91,agranovich97,bamba09crossover}
which is derived from Eq.~\eqref{eq:dispersion}.
If the film surfaces directly contact to external regions,
the Fresnel reflectance coefficients from inside to outside are obtained as
\begin{equation}
r_j(\omega) = \frac{k(\omega) - q_j(\omega)}{k(\omega) + q_j(\omega)}.
\end{equation}
Here, $k(\omega)$ and $q_j(\omega)$
are wavenumbers perpendicular to the surfaces
between the film and external regions ($j = 1, 2$), respectively,
and are defined as
\begin{align}
k(\omega) & = \sqrt{\die(\omega)\omega^2/c^2-\kp{}^2} \label{eq:k(w)}, \\
q_j(\omega) & = \sqrt{\die_j\omega^2/c^2-\kp{}^2},
\end{align}
for wavenumber $\kp$ parallel to the surfaces
and relative dielectric constants $\die_j$ of the two external regions.
The escape rate $\gammabulk(\omega)$ of polariton at frequency $\omega$
is calculated as follows.\cite{agranovich97,bamba09crossover}
After a round trip in the film with a time of $2\thick/\vg(\omega)$,
the density of polaritons decreases by a factor of 
$\exp[-2\gammabulk(\omega)\thick/\vg(\omega)]$,
and it is equal to the factor $|r_{1}(\omega)r_{2}(\omega)|^2$ due to the 
loss at the two surfaces. Then, the escape rate of polariton in a film is 
obtained as
\begin{equation} \label{eq:gamma_escape} 
\gammabulk(\omega)
= \frac{\vg(\omega)}{2\thick} \ln \frac{1}{|r_{1}(\omega)r_{2}(\omega)|^2 }
= \frac{\vg(\omega)}{\thickeff(\omega)},
\end{equation}
where
\begin{equation} \label{eq:Leff} 
\thickeff(\omega) = - \frac{2\thick}{\ln|r_{1}(\omega)r_{2}(\omega)|^2}
\end{equation}
is the effective length for the polariton propagation.
This escape time $\tescape = 1/\gammabulk(\omega)$
reflecting the macroscopic propagation of polaritons
is another time scale in the processes of the P emission.
When the effective thickness is around $\thickeff \sim 5\;\micron$,
the escape time is estimated as $\tescape \sim 0.1\;\ps$
for the P-emission frequency region in ZnO.
\cite{Wakaiki2011pssc,Wakaiki2013EPJB,Wakaiki2014JL}
This is also quite short compared to the observed P-emission lifetime.

\section{Interpretation of P-emission lifetime} \label{sec:P-emission}
\begin{figure}[tbp]
\begin{center}
\includegraphics[width=.9\linewidth]{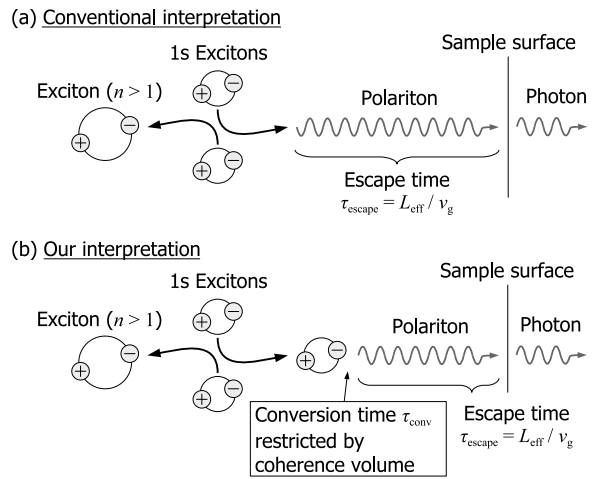}
\caption{Schematic diagrams of (a) conventional interpretation
and (b) our interpretation of the dynamics toward the P emission.
The escape time $\tescape$ of polariton is estimated to be quite short
compared to the observed lifetime of the P emission.
We interpret that the lifetime reflects the conversion time $\tconv$
from scattered excitons to polaritons.
If the excitons after the inelastic scattering
have a coherence length longer than the radiation wavelength,
they can be converted quickly to polaritons as in the conventional interpretation.
However, if the coherence length is quite short,
it restricts the conversion time $\tconv$,
and our interpretation is rather appropriate.}
\label{fig:2}
\end{center}
\end{figure}
Let us consider fundamentally a series of processes
after the inelastic exciton-exciton scattering at the bottleneck region
until photons come out from the sample.
According to the conventional interpretation of the P emission,
as depicted in Fig.~\ref{fig:2}(a),
one of the scattered excitons is converted to a photon-like polariton
almost immediately, because there are only the photon-like polariton states
(eigenstates of electromagnetic fields in medium)
at the P-emission frequency.
In this interpretation,
when polaritons are stabilized by a large enough transition dipole,
they are created in the time scale of the exciton-photon interchange time
$\tRabi = 2\pi/\rabiP_{\mu}$ of the polariton state,
and it is certainly negligible ($\tRabi \sim 0.06\;\ps$ in ZnO)
compared to the other time scales except the escape time $\tescape$ of polariton
(then there is a crossover around the material size comparable to the radiation wavelength
\cite{bamba09crossover}).
Then, if the P-emission lifetimes do not originate from the lifetime
of excitons at the bottleneck,
obeying the conventional interpretation,
we need the interpretations of the polariton diffusion\cite{Hirano2010PRB}
or of the polariton escape from a sample
with an incredibly large effective thickness.\cite{Wakaiki2011pssc,Wakaiki2013EPJB}

Let us examine whether this conventional interpretation
is really justified or not from a fundamental viewpoint.
First of all, even if the polariton states (or photons outside the sample)
are the final states in the processes of the P emission,
we can consider intermediate states between the inelastic scattering
and the escape of polaritons from the sample.
In fact, since the scattering originates from the Coulomb interaction
or the Fermionic nature of excitons,
we originally get two excitons just after the scattering.
The key problem is whether the scattered exciton
is converted to the polariton in the time scale of $\tRabi$ or not.

As discussed in the previous section,
the polariton picture is justified
only when excitons have a long enough spatial coherence,
e.g., when they are created by light irradiation
or after the emission from localized excitons.
In contrast, when the incoherent excitons at the bottleneck
are scattered with each other,
we can consider that the excitons just after the scattering have only a poor spatial coherence.
The conversion from the scattered excitons to polaritons
(or photons outside) is rather similar as the emission process
from localized excitons,
and the conversion time can be restricted by the coherence volume $\volcoh^{\mu}$
of the scattered excitons.

\subsection{Conversion time from exciton to polariton}
Obeying the above scenario, in order to estimate the conversion time
from exciton to polariton,
we need to extend the discussion of
the radiative recombination rate of excitons with the coherence volume
in Sec.~\ref{sec:radiative_decay}.

Although the conversion rate from exciton to photon
is calculated in Sec.~\ref{sec:radiative_decay},
the created photon can be reabsorbed in the excitonic medium
with a large enough size and a large enough transition dipole.
This reabsorption is one of the critical problems
for discussing the emission lifetime of excitons.
Concerning the emission at the bottleneck region,
the created photon is reabsorbed with a relatively high probability,
because the emission frequency is very close to the exciton resonance.
Furthermore, the recreated exciton loses rapidly the memories
of the phase and the propagation direction.
Then, even after the creation of the photon,
it gets hardly a spatial coherence,
and we should consider a repetition of photon creation, reabsorption,
and dephasing of exciton.
This is one of the reasons why the emission lifetime of
bottleneck excitons is hard to be discussed.

On the other hand, the problem can be simplified
when we discuss the P emission.
Since the emission frequency is far below the exciton resonance
(e.g., about $0.1\;\eV$ for ZnO\cite{Wakaiki2011pssc,Wakaiki2013EPJB,Wakaiki2014JL}),
even if the photon is reabsorbed,
the created exciton emits a photon again without losing the memories
of the phase and the propagation direction,
i.e., the absorption coefficient is negligible at that frequency.
Then, after the conversion from a localized exciton to a photon,
we can consider simply the series of absorption and creation of a photon
in the excitonic medium
without the dephasing or scattering process.
In such a case, the propagation of the created photon can be described
by that of the polariton.
Then, instead of the conversion rate from an exciton to a photon
calculated in Sec.~\ref{sec:radiative_decay},
we here calculate the conversion rate from an exciton to a polariton,
because the polariton states are the eigenstates in the medium.

Under the bosonization of the exciton,
the Hamiltonian $\oH_{\text{pol}} = \oH_{\text{rad}} + \oH_{\text{LM}} + \oH_{\text{ex}}$
of photons and excitons can be diagonalized as
\cite{hopfield58,Bamba2013MBC}
\begin{equation}
\oH_{\text{pol}} = \sum_{\eta=1,2} \sum_{j} \sum_{\vk}
\hbar\omega_{j,k} \opd_{j,\vk,\eta} \op_{j,\vk,\eta}.
\end{equation}
Here, $\op_{j,\vk,\eta}$ is the annihilation operator of a polariton
in state $j$ with wavevector $\vk$ and polarization direction $\eta$,
and it is represented approximately for $\rabiP_{\mu,k} \ll \varOmega_{\mu,k}$
by the sum of the annihilation operators of photon and exciton as
\begin{equation}
\op_{j,\vk,\eta} \simeq C_{j,\vk} \oa_{\vk,\eta}
+ \sum_{\mu} X_{j,\vk,\eta,\mu} \osigma_{\mu,\vk}.
\end{equation}
The coefficients are determined by the parameters in the Hamiltonian
$\oH_{\text{pol}} = \oH_{\text{rad}} + \oH_{\text{LM}} + \oH_{\text{ex}}$.
\cite{hopfield58,Bamba2013MBC}
The eigenfrequency of the polariton state is represented as
$\omega_{j,k}$,
and it satisfies $c^2k^2/\omega_{j,k}{}^2 = \die(\omega_{j,k},k)$
for the dielectric function defined in Eq.~\eqref{eq:dispersion}.
For frequency $\omega$ corresponding
to the photon-like region of the lowest polariton branch ($j = L$),
the photonic fraction of the polariton state
is approximately represented by its group velocity $\vg(\omega)$ as
\begin{equation} \label{eq:photonic_fraction_bulk} 
A(\omega) = |C_{L,k_L(\omega)}|^2 \simeq \frac{\vg(\omega)}{\cbg},
\end{equation}
where $k_L(\omega)$ satisfies $c^2k_L(\omega)^2/\omega^2 = \die(\omega,k_L(\omega))$.
Intuitively, if the group velocity is slowed as $\vg(\omega) = A(\omega)\cbg$
compared to the speed $v$ of light in the background medium,
the polariton propagates as a photon in the fraction $A(\omega)$
and as an exciton (its velocity is negligible) in the rest
$1 - A(\omega)$.

After the conversion from the localized exciton to a photon,
it propagates as a polariton stably in the medium
in the P-emission frequency region,
and the interchange rate $\rabiP_{\mu,k}$ is much higher than
the conversion rate $\varGamma_{\mu}$.
Then, before the escape of the polariton from the sample,
the final state can be supposed as the lowest polariton state $\opd_{L,\vk,\eta}\ket{0}$.
The initial state is the exciton after the inelastic scattering,
and here it is supposed as a mixed one concerning several exciton's relative-motion states
represented in Eq.~\eqref{eq|ex_mu>},
i.e., the density operator of one exciton state is expressed as
\begin{equation} \label{eq:|ex>} 
\hat{\rho}_{\text{init}}^{\text{one}}
= \sum_{\mu} f_{\mu} \ket{\text{ex}_{\mu}}\bra{\text{ex}_{\mu}},
\end{equation}
where $f_{\mu}$ represents the probability
of being in the state $\mu$ ($\sum_{\mu}f_{\mu} = 1$).
Instead of Eq.~\eqref{eq:Gammma_mu},
the conversion rate is derived as
\begin{align}
\varGamma(\omega)
& = \frac{2\pi}{\hbar}\sum_{\eta,\vk,\mu} f_{\mu}
    |\braket{0|\op_{L,\vk,\eta}\oH_{\text{LM}}|\text{ex}_{\mu}}|^2
    \delta(\hbar\omega-\hbar\omega_{L,k}) \nonumber \\
& = \frac{2\pi}{\hbar}\sum_{\eta,\vk,\mu}f_{\mu}
    |C_{L,k}|^2
    |\braket{0|\oa_{\vk,\eta}\oH_{\text{LM}}|\text{ex}_{\mu}}|^2
    \delta(\hbar\omega-\hbar\omega_{L,k}) \nonumber \\
& = A(\omega) \sum_{\mu} f_{\mu} \gamma_{\mu} \frac{\volcoh^{\mu}}{\volunit}
    \left[ \frac{vk_L(\omega)}{\varOmega_{\mu,k_L(\omega)}} \right]^3.
\label{eq:Gamma(w)} 
\end{align}
Through the factor $A(\omega)$,
this conversion rate includes the contribution of reabsorption and reemission of the photon
but without the dephasing or the scattering process.
Note that,
whereas the first line is derived from the Fermi's golden rule,
the energy $\hbar\omega$ corresponds to the eigenenergy
of the final state $\opd_{L,\vk_L(\omega),\eta}\ket{0}$
but not the energy of the initial state $\hat{\rho}_{\text{init}}^{\text{one}}$
or $\ket{\text{ex}_{\mu}}$,
which is far above the emission energy $\hbar\omega$.
This point will be discussed in Sec.~\ref{sec:discussion}.

The $\omega$-dependence comes from the two factors:
$A(\omega)\propto \vg(\omega)$
and $[\cbg k_L(\omega)/\varOmega_{\mu,k_L(\omega)}]^3$.
Around the P-emission frequency in most of the materials,
the former gives the dominant contribution than the latter,
which is almost unity and gives a slight $\omega$-dependence.
For example, in ZnO, we have $\hbar\Omega_{A,1s} = 3.375\;\eV$ and
$\hbar\omega \sim \hbar\Omega_{A,1s} - 0.1\;\eV$.
Since the frequency difference $0.1\;\eV$ is in the same order as
$\hbar\rabiP_{A,1s} = 70\;\meV$,
the photonic fraction $A(\omega) \propto \vg(\omega)$ gives the dominant contribution
around this frequency region,
and the conversion rate can be approximately expressed as
\begin{equation}
\varGamma(\omega) \simeq A(\omega)
  \sum_{\mu} f_{\mu} \gamma_{\mu} \frac{\volcoh^{\mu}}{\volunit}.
\end{equation}
This can be the reason why the P-emission lifetime is observed
to be inversely proportional to the group velocity $\vg(\omega)$
in the experiments for bulk materials.
\cite{Wakaiki2011pssc,Wakaiki2013EPJB,Wakaiki2014JL}
Whereas the escape time $\tescape$ of polaritons after the conversion
is also inversely proportional to the group velocity $\vg(\omega)$,
it is estimated to be quite short
compared to the conversion time $\tconv = 1/\varGamma(\omega)$.
In this way, 
the P-emission lifetime basically reflects the conversion time $\tconv$
from the scattered excitons to the polaritons
in our interpretation.

\subsection{Estimation of coherence volume} \label{sec:estimate_Vcoh}
From the experimental data for ZnO,\cite{Wakaiki2014JL,Wakaiki2015PE}
we here estimate the coherence volume of the scattered excitons
based on the expression \eqref{eq:Gamma(w)}
of the exciton-to-polariton conversion rate $\varGamma(\omega)$.
Since the A and B excitons are the lowest two states,
here we tentatively consider that these two exciton states are mostly created
at the P-emission frequency region by the inelastic scattering,
i.e., $f_{A,1s} + f_{B,1s} = 1$.
Further, the radiative recombination rates for exciton localized in a unit cell
are similar $\gamma_{A,1s} \sim \gamma_{B,1s}$
for the two exciton states,
because we have $\rabiP_{A,1s} \sim \rabiP_{B,1s}$ as discussed in
Sec.~\ref{eq:interchange}.
The eigenfrequencies are also similar as $\varOmega_{A,1s} \sim \varOmega_{B,1s}$.
Then, the conversion rate is rewritten approximately as
\begin{equation}
\varGamma(\omega)
= A(\omega)
  \left[ \frac{vk_L(\omega)}{\varOmega_{A,1s,k_L(\omega)}} \right]^3
  \varGamma',
\end{equation}
where the $\omega$-independent decay rate is defined
with an averaged coherence volume
$\volcoh = f_{A,1s}\volcoh^{A,1s} + f_{B,1s}\volcoh^{B,1s}$ as
\begin{equation}
\varGamma' = \gamma_{A,1s} \frac{\volcoh}{\volunit}.
\end{equation}
From the experimentally obtained P-emission lifetimes,
we estimated $\varGamma' = (0.8\;\ps)^{-1}$.
Then, from the radiative recombination rate $\gamma_{A,1s} = 0.45\;(\mu\sec)^{-1}$
derived for an exciton localized at a unit cell
in Eq.~\eqref{eq:gamma_A1s},
the coherence volume is estimated as
\begin{equation}
\volcoh = 6\times10^7(\AA)^3,
\end{equation}
and the coherence length is $(\volcoh)^{1/3} = 4\times10^1\;\nm$.
Although we have currently no other way to evaluate the coherence volume (length)
experimentally,
this value is certainly shorter than the radiation wavelength
($\sim 2\times10^2\;\nm$ for $\hbar\omega = 3.26\;\eV$ in the background medium with $\diebg = 4$).

In this way, from the fundamental viewpoint,
we should consider the coherence volume of the scattered excitons,
and the conversion time from the exciton to the photon-like polariton
can explain the observed P-emission lifetime,
which is much shorter than the emission lifetime $\temit$ of the bottleneck excitons,
longer than the escape time $\tescape$ of polaritons,
and inversely proportional to the group velocity approximately for bulk materials.
Whereas the discussion in this paper does not deny the interpretation
of the polariton diffusion,\cite{Hirano2010PRB}
it is note that the decrease in diffusion constant
with an increase in impurity concentration reported in Ref.~\onlinecite{Hirano2010PRB}
can be explained as a decrease in coherence volume $\volcoh$ in our interpretation.

In the next section, we try to justify our interpretation
against some counter-intuitive points.

\section{Discussion} \label{sec:discussion}
Since the final states certainly exist as the polariton states
or photon states outside the sample,
the inelastic scattering to these destinations is not forbidden.
However, the scattered excitons remain in the bare exciton states
in the conversion time $\tconv \sim 1\;\ps$,
although the eigenfrequencies $\wx_{\mu,k}$ of these states
are far above the emission frequency $\omega$
($\hbar\wx_{A,1s} = 3.375\;\eV$ and $\hbar\wx_{A,1s}-\hbar\omega \sim 0.1\;\eV$
for ZnO\cite{Wakaiki2014JL,Wakaiki2015PE}).
In the conventional interpretation,
the inelastic scattering of the two excitons is resonant to
both the higher exciton state with $n>1$ and the photon-like polariton one.
In contrast, in our interpretation,
it is resonant only to the higher exciton state
but not to the lower one (no exciton state at the P-emission frequency).
However, even if one process is not resonant,
it can occur in a series of processes.

Since the A and B exciton states with $n = 1$ are most resonant
compared to the other exciton states
($\wx_{A,1s}$ and $\wx_{B,1s}$ are closest to the emission frequency $\omega$),
the scattered excitons are supposed to be mostly
in the lowest two exciton states
($f_{A,1s} + f_{B,1s} \sim 1$ and $f_{A,1s} > f_{B,1s}$).
\footnote{
In our interpretation, since the exciton is scattered not directly to the photon-like polariton state,
dark exciton states can also be the transient state
if the final destination exists.
However, we cannot say anything about the dark excitons
from the current experiments of the P emission.
}
These facts justify the estimation of the coherence volume
$\volcoh$ in Sec.~\ref{sec:estimate_Vcoh}.

In our interpretation, the scattered excitons remain in the bare exciton states
not as the so-called virtual state, whose lifetime is determined
by the Heisenberg uncertainty principle \cite{Boitier2009NP}
such as $2\pi/(\wx_{A,1s}-\omega) \sim 0.04\;\ps \ll \tconv$ in our case.
The P-emission process can be a good example
for investigating the validity conditions of the virtual-state picture.
From the experimentally observed lifetime,
we conclude that the virtual-state picture is not appropriate
for the P emission process.
In order to investigate theoretically the validity conditions,
we must consider explicitly the series of the processes
including the inelastic scattering
as will be discussed in Sec.~\ref{sec:micro}.

Intuitively,
the bottleneck excitons are scattered
to unstable transient states (bare exciton states) with a lifetime of $\tconv$.
If the dephasing time of the higher excitons ($n > 1$)
is shorter than $\tconv$, the emission frequency $\omega$ is fixed
during the excitons remain in the transient states.
Such transient states are surely unstable,
and then $\tconv$ is much shorter than the emission lifetime $\temit$
of excitons at the bottleneck region.
The conversion time $\tconv$ becomes shorter (less stable)
with a decrease in the emission frequency
(more distant from the bottleneck frequency).

The conversion rate $\varGamma(\omega)$ from exciton to polariton
was calculated in Eq.~\eqref{eq:Gamma(w)}.
In this derivation,
the emission energy $\hbar\omega$ corresponds to the energy of the final state (polariton)
but is lower than that of the initial state $\hat{\rho}_{\text{init}}^{\text{one}}$,
which is at least higher than the lowest exciton state
$\Tr(\oH_{\text{ex}}\hat{\rho}_{\text{init}}^{\text{one}}) > \hbar\wx_{A,1s,k=0}$.
On the other hand, 
in the spontaneous emission from the bottleneck excitons,
we can suppose $\hbar\omega \sim \hbar\wx_{\mu}$,
and the exciton state $\ket{\text{ex}_{\mu}}$
can be supposed well as an initial state.
In order to describe more rigorously the transient state in the P emission process,
we try to discuss the series of processes including the inelastic processes
in the followings.

\subsection{Inelastic exciton-exciton scattering} \label{sec:micro}
In order to examine strictly whether the lifetime of the transient state
$\ket{\text{ex}_{\mu}}$ is really restricted by the coherence volume,
instead of starting from the scattered excitons as in the previous section,
we need to consider the series of processes of the P emission
from the inelastic scattering to the creation of the polaritons.
Here, we treat the exciton operator $\osigma_{\mu,\vk}$ as bosonic one
and suppose the Hamiltonian of the exciton-exciton interaction as
\begin{equation}
\oH_{\text{ex-ex}} = \sum_{\mu,\nu,\mu',\nu'} \sum_{\vk,\vk',\vq}
\frac{\hbar\sct_{\mu,\mu',\nu',\nu,\vq}}{2}
\osigmad_{\mu,\vk}\osigmad_{\mu',\vk'}\osigma_{\nu',\vk'-\vq}\osigma_{\nu,\vk+\vq},
\end{equation}
where the scattering coefficient satisfies
$\sct_{\mu,\mu',\nu',\nu,\vq} = \sct_{\mu,\mu',\nu',\nu,-\vq}
= \sct_{\mu',\mu,\nu',\nu,\vq} = \sct_{\nu,\nu',\mu',\mu,\vq}$.
From the Hamiltonian $\oH_{\text{rad}} + \oH_{\text{LM}} + \oH_{\text{ex}} + \oH_{\text{ex-ex}}$,
the Heisenberg equations of $\oa_{\vk,\eta}$ and $\osigma_{\mu,\vk}$ are derived
under the rotating-wave approximation as
\begin{subequations} \label{eq:Heisenberg_ex-ex} 
\begin{align}
\ii\ddt{}\oa_{\vk,\eta}
& = \cbg|\vk|\oa_{\vk,\eta}
  + \sum_{\mu}\ii\rabiP_{\mu,\vk,\eta} \osigma_{\mu,\vk}, \\
\ii\ddt{}\osigma_{\mu,\vk}
& = \wx_{\mu,k}\osigma_{\mu,\vk}
  - \sum_{\eta}\ii\rabiP_{\mu,\vk,\eta}\oa_{\vk,\eta}
\nonumber \\ & \quad
  + \sum_{\nu,\mu',\nu'} \sum_{\vk',\vq} \sct_{\mu,\mu',\nu',\nu,\vq}
    \osigmad_{\mu',\vk'} \osigma_{\nu',\vk'-\vq} \osigma_{\nu,\vk+\vq},
\end{align}
\end{subequations}
where $\rabiP_{\mu,\vk,\eta} = \vunit_{\mu}\cdot\vunit_{\vk,\eta}\rabiP_{\mu,k}$.

We also suppose that the excitonic system is in a quasi-equilibrium
at the bottleneck region, and the one-exciton density operator is
represented as
\begin{equation} \label{eq:rho_one_exciton} 
\orho_{\text{eq}}^{\text{one}}
= \sum_{\mu,\lambda} \frac{P_{\mu}}{N}
  \ket{\text{ex}_{\mu,\lambda}} \bra{\text{ex}_{\mu,\lambda}}
= \sum_{\mu,\vk} P_{\mu}|\varPhi_{\mu,\vk}|^2 \osigmad_{\mu,\vk}\ket{0}\bra{0}\osigma_{\mu,\vk}.
\end{equation}
Here, $P_{\mu}$ is the probability of being in the $\mu$ state.
The exciton exists anywhere in the whole space with an equal probability.
The center-of-mass of exciton spreads coherently in space with the wavefunction $\varPhi_{\mu}(\vr)$,
and the exciton state located around $\vR_{\lambda}$ is represented as
\begin{subequations}
\begin{align}
\ket{\text{ex}_{\mu,\lambda}}
& = \sum_{\lambda'} \sqrt{\volunit} \varPhi_{\mu}(\vR_{\lambda'}-\vR_{\lambda})
    \osigmad_{\mu,\lambda'}\ket{0}, \\
& = \sum_{\vk} \ee^{-\ii\vk\cdot\vR_{\lambda}} \varPhi_{\mu,\vk}
    \osigmad_{\mu,\vk}\ket{0},
\end{align}
\end{subequations}
where the Fourier transform of $\varPhi_{\mu}(\vr)$ is defined as
\begin{equation}
\varPhi_{\mu,\vk} = \frac{1}{\sqrt{\vol}}
\int\dd\vr\ \ee^{-\ii\vk\cdot\vr}\varPhi_{\mu}(\vr).
\end{equation}
Eq.~\eqref{eq:rho_one_exciton} means that
the exciton spreads in the $k$-space
not as the distribution of many excitons
but as the one-exciton mixed state
reflecting the finiteness of the coherence volume.
From this one-exciton density operator,
we get the following expectation values
\begin{subequations} \label{eq:one-body_correlation} 
\begin{align}
\braket{\osigmad_{\mu,\lambda}\osigma_{\mu',\lambda'}}_{\text{eq}}^{\text{one}}
& = \delta_{\mu,\mu'}\frac{P_{\mu}}{N}
    \int\dd\vr\ \varPhi_{\mu}(\vR_{\lambda}-\vr)^*\varPhi_{\mu}(\vR_{\lambda'}-\vr), \\
\braket{\osigmad_{\mu,\vk}\osigma_{\mu',\vk'}}_{\text{eq}}^{\text{one}}
& = \delta_{\mu,\mu'}\delta_{\vk,\vk'}
    P_{\mu} |\varPhi_{\mu,\vk}|^2.
\end{align}
\end{subequations}
In the followings, we suppose that
each exciton shows this spatial distribution in the quasi-equilibrium state
consisting of many excitons,
and the one-body correlation in the quasi-equilibrium is written as
\begin{equation} \label{eq:<sds>_k} 
\braket{\osigmad_{\mu,\vk}\osigma_{\mu',\vk'}}_{\text{eq}}
= \delta_{\mu,\mu'}\delta_{\vk,\vk'} N_{\mu,\vk},
\end{equation}
where $N_{\mu,\vk} \propto P_{\mu}|\varPhi_{\mu,\vk}|^2$
represents the expectation number of excitons
in state $\mu$ and with wavevector $\vk$.

Let us discuss the inelastic exciton-exciton scattering process
as a perturbation to the quasi-equilibrium state.
In the Heisenberg picture, the equations of
deviation operators $\delta\oa_{\vk,\eta} = \oa_{\vk,\eta} - \oa_{\vk,\eta}^{\text{eq}}$
and $\delta\osigma_{\mu,\vk} = \osigma_{\mu,\vk} - \osigma_{\mu,\vk}^{\text{eq}}$
from the quasi-equilibrium are obtained from Eqs.~\eqref{eq:Heisenberg_ex-ex} as
\begin{widetext}
\begin{subequations}
\begin{align}
\ii\ddt{}\delta\oa_{\vk,\eta}
& \simeq \cbg|\vk|\delta\oa_{\vk,\eta}
  + \sum_{\mu} \ii\rabiP_{\mu,\vk,\eta} \delta\osigma_{\mu,\vk},
\label{eq:doa_k} \\ 
\ii\ddt{}\delta\osigma_{1s,\xi,\vk}
& \simeq \wx_{1s,k}\delta\osigma_{1s,\xi,\vk}
  - \sum_{\eta}\ii\rabiP_{1s,\xi,\vk,\eta}\delta\oa_{\vk,\eta}
  + \sum_{\mu\neq1s} \sum_{\xi',\xi''} \sum_{\vk',\vq}
    \left[
      \sct_{\mu,\xi,\xi',\xi'',\vq}
      \delta(\osigmad_{\mu,\vk'} \osigma_{1s,\xi',\vk'-\vq} \osigma_{1s,\xi'',\vk+\vq})
\right. \nonumber \\ & \quad \left.
    + \sct_{\mu,\xi'',\xi',\xi,\vq}
      \delta(\osigmad_{1s,\xi',\vk'} \osigma_{1s,\xi'',\vk'-\vq} \osigma_{\mu,\vk+\vq})
    \right],
\label{eq:dosigma_1s} \\ 
\ii\ddt{}\delta\osigma_{\mu,\vk}
& \simeq \wx_{\mu,k}\delta\osigma_{\mu,\vk}
  - \sum_{\eta}\ii\rabiP_{\mu,\vk,\eta}\delta\oa_{\vk,\eta}
  + \sum_{\xi,\xi',\xi''} \sum_{\vk',\vq} \sct_{\mu,\xi,\xi',\xi'',\vq}
    \delta(\osigmad_{1s,\xi,\vk'} \osigma_{1s,\xi',\vk'-\vq} \osigma_{1s,\xi'',\vk+\vq}).
\label{eq:dosigma_mu} 
\end{align}
\end{subequations}
Here, we supposed simply that
the quasi-equilibrium state consists of only the lowest excitons (1s)
and the 1s exciton state has only the degrees of freedom of polarization direction $\xi = \{x,y,z\}$
and of wavevector $\vk$.
In the above equations, we keep only the terms involving
the inelastic exciton-exciton scattering process.
The third term in Eq.~\eqref{eq:dosigma_1s} represents the scattering
from two 1s excitons to $\mu\neq 1s$ and 1s excitons,
and the last term represents the inverse process.
The last term in Eq.~\eqref{eq:dosigma_mu} also represents the former process.
When we consider that Eq.~\eqref{eq:dosigma_1s}
describes the development of scattered excitons converting to the photon-like polaritons,
the last two terms in it represent the creation
of the scattered excitons in the $(1s,\xi,\vk)$ state.
In this case, the most important term is the third term,
and the deviation operator in it is expanded up to the lowest order as
\begin{equation} \label{eq:d(sss)} 
\delta(\osigmad_{\mu,\vk'} \osigma_{1s,\xi',\vk'-\vq} \osigma_{1s,\xi'',\vk+\vq})
\simeq 
\delta\osigmad_{\mu,\vk'} \osigma^{\text{eq}}_{1s,\xi',\vk'-\vq} \osigma^{\text{eq}}_{1s,\xi'',\vk+\vq}
+ \osigma^{\text{eq}\dagger}_{\mu,\vk'} \delta\osigma_{1s,\xi',\vk'-\vq} \osigma^{\text{eq}}_{1s,\xi'',\vk+\vq}
+ \osigma^{\text{eq}\dagger}_{\mu,\vk'} \osigma^{\text{eq}}_{1s,\xi',\vk'-\vq} \delta\osigma_{1s,\xi'',\vk+\vq}.
\end{equation}
These $(1s,\xi',\vk'-\vq)$ and $(1s,\xi'',\vk+\vq)$ states
correspond to the two bottleneck excitons,
and $(\mu,\vk')$ is the higher exciton state.
For the unperturbed time-development of these three states,
we can neglect the coupling with photons.
Then, from Eqs.~\eqref{eq:dosigma_1s} and \eqref{eq:dosigma_mu},
the deviation operators on the right-hand side in Eq.~\eqref{eq:d(sss)}
are expressed approximately as
\begin{subequations}
\begin{align}
\delta\osigma_{1s,\xi,\vk}(t)
& \simeq -\ii\int_{t_0}^{t}\dd t'\ \ee^{-\ii\wx_{1s,k}(t-t')}
    \sum_{\mu\neq1s} \sum_{\xi',\xi''} \sum_{\vk',\vq}
    \left[
      \sct_{\mu,\xi,\xi',\xi'',\vq}
      \delta(\osigmad_{\mu,\vk'} \osigma_{1s,\xi',\vk'-\vq} \osigma_{1s,\xi'',\vk+\vq})(t')
\right. \nonumber \\ & \quad \left.
    + \sct_{\mu,\xi'',\xi',\xi,\vq}
      \delta(\osigmad_{1s,\xi',\vk'} \osigma_{1s,\xi'',\vk'-\vq} \osigma_{\mu,\vk+\vq})(t')
    \right], \\
\delta\osigmad_{\mu,\vk}(t)
& \simeq \ii\int_{t_0}^{t}\dd t'\ \ee^{\ii\wx_{\mu,k}(t-t')}
    \sum_{\xi,\xi',\xi''} \sum_{\vk',\vq} \sct_{\mu,\xi,\xi',\xi'',\vq}
    \delta(\osigmad_{1s,\xi'',\vk+\vq} \osigmad_{1s,\xi',\vk'-\vq} \osigma_{1s,\xi,\vk'})(t'),
\end{align}
\end{subequations}
where $t_0$ is the starting time of the inelastic scattering process.
We substitute these into the third term of Eq.~\eqref{eq:dosigma_1s}
through the expansion \eqref{eq:d(sss)},
and we neglect the fourth term, i.e., the inverse process.
Further, we keep only the terms proportional to $\delta\osigma_{1s,\xi,\vk}$,
which are the dominant terms
because they involve the stimulated emission of polaritons
or stimulated creation of excitons (this point will be discussed in the next subsection).
Finally linearizing the equation with respect to the deviation operator, we get
\begin{align}
\ii\ddt{}\delta\osigma_{1s,\xi,\vk}(t)
& \simeq \wx_{1s,k}\delta\osigma_{1s,\xi,\vk}(t)
  - \sum_{\eta}\ii\rabiP_{1s,\xi,\vk,\eta}\delta\oa_{\vk,\eta}(t)
  + \sum_{\mu\neq1s} \sum_{\xi',\xi''} \sum_{\vk',\vq} \ii\sct_{\mu,\xi,\xi',\xi'',\vq}\int_{t_0}^{t}\dd t'
    \sum_{\zeta,\zeta'} \sum_{\vk'',\vq'}
    [
\nonumber \\ & \quad
     \ee^{\ii\wx_{\mu,\vk'}(t-t')}
     \sum_{\zeta''}\sct_{\mu,\zeta,\zeta',\zeta'',\vq'}
     \braket{\osigmad_{1s,\zeta'',\vk'+\vq'}(t') \osigmad_{1s,\zeta',\vk''-\vq'}(t')
     \osigma_{1s,\xi',\vk'-\vq}(t) \osigma_{1s,\xi'',\vk+\vq}(t)}_{\text{eq}}
     \delta\osigma_{1s,\zeta,\vk''}(t')
\nonumber \\ & \quad
    - \ee^{-\ii\wx_{1s,\vk'-\vq}(t-t')}
      \sct_{\mu,\zeta,\zeta',\xi',\vq'}
      \braket{\osigmad_{\mu,\vk'}(t)\osigmad_{1s,\zeta',\vk''}(t')
      \osigma_{\mu,\vk'-\vq+\vq'}(t')\osigma_{1s,\xi'',\vk+\vq}(t)}_{\text{eq}}
      \delta\osigma_{1s,\zeta,\vk''-\vq'}(t')
\nonumber \\ & \quad
    - \ee^{-\ii\wx_{1s,\vk+\vq}(t-t')}
      \sct_{\mu,\zeta,\zeta',\xi'',\vq'}
      \braket{\osigmad_{\mu,\vk'}(t) \osigma_{1s,\xi',\vk'-\vq}(t)
      \osigmad_{1s,\zeta',\vk''}(t')\osigma_{\mu,\vk+\vq+\vq'}(t')}_{\text{eq}}
      \delta\osigma_{1s,\zeta,\vk''-\vq'}(t')
    ].
\label{eq:ddt_ds_1} 
\end{align}
In principle, the two-body correlation functions in the quasi-equilibrium
must be calculated by considering the elastic exciton-exciton scattering process,
interaction with phonons, radiative recombination of excitons, etc.
However, here we approximate them simply by products
of the one-body correlations given in Eq.~\eqref{eq:<sds>_k}
with introducing phenomenologically dephasing rates $\gammadeph_{\mu,\vk}$ as
\begin{subequations}
\begin{align}
\braket{\osigmad_{1s,\zeta'',\vk'+\vq'}(t') \osigmad_{1s,\zeta',\vk''-\vq'}(t')
        \osigma_{1s,\xi',\vk'-\vq}(t) \osigma_{1s,\xi'',\vk+\vq}(t)}_{\text{eq}}
& = \delta_{\vk'',\vk}\delta_{\vq',-\vq}\delta_{\zeta',\xi''}\delta_{\zeta'',\xi'}
    N_{1s,\xi',\vk'-\vq}(t')N_{1s,\xi'',\vk+\vq}(t')
\nonumber \\ & \quad \times
    \ee^{[-\ii(\wx_{1s,\vk+\vq}+\wx_{1s,\vk'-\vq})-(\gammadeph_{1s,\vk+\vq}+\gammadeph_{1s,\vk'-\vq})](t-t')}, \\
\braket{\osigmad_{\mu,\vk'}(t)\osigmad_{1s,\zeta',\vk''}(t')
        \osigma_{\mu,\vk'-\vq+\vq'}(t')\osigma_{1s,\xi'',\vk+\vq}(t)}_{\text{eq}}
& = \delta_{\vk'',\vk+\vq} \delta_{\vq',\vq} \delta_{\zeta',\xi''}
    N_{1s,\xi'',\vk+\vq}(t') N_{\mu,\vk'}(t')
\nonumber \\ & \quad \times
    \ee^{[-\ii(\wx_{1s,\vk+\vq}-\wx_{\mu,\vk'}))-(\gammadeph_{1s,\vk+\vq}+\gammadeph_{\mu,\vk'})](t-t')}, \\
\braket{\osigmad_{\mu,\vk'}(t) \osigma_{1s,\xi',\vk'-\vq}(t)
        \osigmad_{1s,\zeta',\vk''}(t')\osigma_{\mu,\vk+\vq+\vq'}(t')}_{\text{eq}}
& = \delta_{\vk'',\vk'-\vq}\delta_{\vq',\vk'-\vk-\vq}\delta_{\zeta',\xi'}
    (1+N_{1s,\xi',\vk'-\vq}(t'))N_{\mu,\vk'}(t')
\nonumber \\ & \quad \times
    \ee^{[-\ii(\wx_{1s,\vk'-\vq}-\wx_{\mu,\vk'}))-(\gammadeph_{1s,\vk'-\vq}+\gammadeph_{\mu,\vk'})](t-t')}.
\end{align}
\end{subequations}
Here, the density $N_{\mu,\vk}(t)$ of excitons is supposed to depend on time.
Under this approximation,
keeping only the dominant terms involving the stimulated process,
Eq.~\eqref{eq:ddt_ds_1} is rewritten as
\begin{equation}
\ii\ddt{}\delta\osigma_{1s,\xi,\vk}(t)
\simeq \wx_{1s,k}\delta\osigma_{1s,\xi,\vk}(t)
  - \sum_{\eta}\ii\rabiP_{1s,\xi,\vk,\eta}\delta\oa_{\vk,\eta}(t)
  + \int_{t_0}^t\dd t' \int\dd\omega\ \ii S_{\xi,\vk}(\omega,t,t') \ee^{-\ii\omega(t-t')}
    \delta\osigma_{1s,\xi,\vk}(t'),
\label{eq:ddt_ds_2} 
\end{equation}
where the integral kernel is expressed as
\begin{align}
S_{\xi,\vk}(\omega,t,t')
& = \sum_{\mu\neq1s} \sum_{\xi',\xi''} \sum_{\vk',\vq}
    \delta(\omega+\wx_{\mu,\vk'}-\wx_{1s,\vk+\vq}-\wx_{1s,\vk'-\vq})
    \sct_{\mu,\xi,\xi',\xi'',\vq}{}^2
\nonumber \\ & \quad \times
    \left[ N_{1s,\xi',\vk'-\vq}(t')N_{1s,\xi'',\vk+\vq}(t')
      \ee^{-(\gammadeph_{1s,\vk+\vq}+\gammadeph_{1s,\vk'-\vq})(t-t')}
\right. \nonumber \\ & \quad \left.
    - (1+2N_{1s,\xi',\vk+\vq}(t'))N_{\mu,\vk'}(t')
    \ee^{-(\gammadeph_{1s,\vk+\vq}+\gammadeph_{\mu,\vk'})(t-t')}
    \right].
\label{eq:S_k(w,t)} 
\end{align}
In this way, under the above approximations,
the equation is reduced to the one-body one consisting of
Eqs.~\eqref{eq:doa_k} and \eqref{eq:ddt_ds_2}.
What we have to solve is the master equation derived from these equations as
\begin{equation} \label{eq:master} 
\ddt{}\orho(t) = \frac{1}{\ii\hbar}\left[ \oH_{\text{pol}}, \orho(t) \right]
+ \mathcal{L}_{\text{diss}}[\orho]
+ \sum_{\vk}\int\dd\omega\int_{t_0}^t\dd t'\ S_{\vk}(\omega,t,t') \left\{
    \ee^{\ii\omega(t-t')}\left[ \osigmad_{\vk}\orho(t'), \osigma_{\vk} \right]
  + \ee^{-\ii\omega(t-t')}\left[ \osigmad_{\vk}, \orho(t')\osigma_{\vk} \right]
  \right\}.
\end{equation}
\end{widetext}
The second term $\mathcal{L}_{\text{diss}}[\orho]$
is introduced for the dissipation of excitons and photons.
The last two terms come from the last term in Eq.~\eqref{eq:ddt_ds_2}.
They originate from the inelastic exciton-exciton scattering,
and gives a gain for the creation of excitons or polaritons.
Due to the presence of these terms,
the P emission shows a threshold behavior
involving the stimulated emission of polaritons
or stimulated creation of excitons.
The decay of the exciton density $N_{\mu,\vk}(t)$ appearing in Eq.~\eqref{eq:S_k(w,t)}
should be solved together with Eq.~\eqref{eq:master}.
This problem will be discussed in the next subsection.
Once we obtain the temporal evolution of $N_{\mu,\vk}(t)$,
we can calculate the correlation of polaritons
$\braket{\opd_{j,\vk,\eta}(t')\op_{j,\vk,\eta}(t)}$,
which gives the P-emission spectra
and the exciton-to-polariton conversion time.

Since the polariton states are the eigenstates of the unperturbed Hamiltonian
$\oH_{\text{pol}}$,
the deviation operators can be approximated as $\delta\osigma_{1s,\xi,\vk}(t)
\simeq \sum_{j,\eta} X_{j,\vk,\eta,1s,\xi}\delta\op_{j,\vk,\eta}\ee^{-\ii\omega_{j,k}t}$.
Substituting this into the last term in Eq.~\eqref{eq:ddt_ds_2},
we get the quasi-conservation of the energy as
$\omega_{j,\vk} + \wx_{\mu,\vk'} \sim \wx_{1s,\vk+\vq} + \wx_{1s,\vk'-\vq}$,
if the dephasing rate $\gammadeph_{\mu,k}$ is low enough compared to the oscillation frequency
$\omega_{j,\vk}$.
Only the lowest polariton ($j=L$) can satisfy this energy conservation,
and its density should be finally enhanced compared with the other polariton states.
However, it is dangerous to approximate
$\delta\osigma_{1s,\xi,\vk}(t)
\simeq X_{L,\vk,\eta,1s,\xi}\delta\op_{L,\vk,\eta}\ee^{-\ii\omega_{L,k}t}$,
because the exciton-to-polariton conversion can be restricted
by the coherence volume of the scattered excitons
as discussed in the previous section.
We should solve the master equation \eqref{eq:master}
in the photon-exciton basis
or in the basis consisting of all the polariton states
in principle.
In the conventional interpretation, the energy conservation
determines the wavevector of the scattered state
(destination of the inelastic scattering).
On the other hand, in our interpretation,
the scattered excitons spread in the $k$-space reflecting the coherence volume,
and the exciton states are forced to oscillate with the frequency $\omega$
as seen in Eqs.~\eqref{eq:ddt_ds_2} and \eqref{eq:master}.
In both interpretation, the $\omega$-dependence of $S_{\vk}(\omega,t,t')$
basically gives the P-emission spectra,
and it is determined
by the $k$-distribution of the bottleneck excitons $N_{1s,\xi,\vk}$
and of $\sct_{\mu,\xi,\xi',\xi'',\vk}$
through $S_{\vk}(\omega,t,t')$ in Eq.~\eqref{eq:S_k(w,t)}.

Although the wavevector $\vk$ seems to be a good quantum number
in Eqs.~\eqref{eq:doa_k} and \eqref{eq:ddt_ds_2},
all the $\vk$-components are in fact connected in the master equation
\eqref{eq:master}.
Whereas we should in principle calculate $\rho(t)$ and $N_{\mu,\vk}(t)$ self-consistently,
we here focus on only the exciton-to-polariton conversion process.
Such a situation can be considered by assuming
$N_{1s,\xi',\vk'-\vq}(t)N_{1s,\xi'',\vk+\vq}(t)
= \delta(t)N_{1s,\xi',\vk'-\vq}(0)N_{1s,\xi'',\vk+\vq}(0)$
in Eq.~\eqref{eq:S_k(w,t)} and $t_0 < 0$.
For simplicity, we also suppose $N_{\mu\neq1s,\vk} = 0$,
i.e., the density of the higher excitons is negligible
compared to that of the bottleneck excitons.
Under these approximations, the master equation \eqref{eq:master}
is simplified as
\begin{align}
\ddt{}\orho(t)
& = \frac{1}{\ii\hbar}\left[ \oH_{\text{pol}}, \orho(t) \right]
+ \int\dd\omega \sum_{\vk}\left\{
  \ee^{-\ii\omega t} S_{\vk}(\omega,t,0)
\right. \nonumber \\ & \quad \times \left.
    \left[ \osigmad_{1s,\xi,\vk}, \orho(0)\osigma_{1s,\xi,\vk}\right]
  + \Hc \right\}.
\label{eq:master_simple} 
\end{align}
The last terms create excitons during the dephasing time
$1/(\gammadeph_{1s,\vk+\vq}+\gammadeph_{1s,\vk'-\vq})$,
which should be long enough than the oscillation period $2\pi/\omega$.
After that, the created exciton is converted to a polariton
as an one-body problem in the Hamiltonian $\oH_{\text{pol}}$,
in which the frequency mixing (nonlinear process) does not occur.
Then, the problem can be reduced to the exciton-to-polariton conversion
as discussed in the previous section.
The initial exciton state
defined in Eq.~\eqref{eq|ex_mu>} or Eq.~\eqref{eq:|ex>}
is determined by the last two terms in Eq.~\eqref{eq:master_simple},
and the center-of-mass motion of excitons
is distributed for emission frequency $\omega$ as
\begin{equation}
|\varPhi_{1s,\xi,\vk}|^2
\propto S_{\xi,\vk}(\omega,0,0).
\end{equation}
In this way, the coherence volume $\volcoh$ of the scattered excitons
is determined mainly
by the $k$-distributions of density $N_{1s,\xi,\vk}(t)$ of bottleneck excitons
and of scattering coefficient $\sct_{\mu,\xi,\xi',\xi'',\vk}$
through Eq.~\eqref{eq:S_k(w,t)}.
In contrast to the interpretation of the direct creation of polaritons,
the delta function in Eq.~\eqref{eq:S_k(w,t)} (energy conservation)
does not determine the wavevector of the scattered excitons,
and its center-of-mass wavefunction $|\varPhi_{1s,\xi,\vk}|^2$ spreads
in the region of $|\vk| \lesssim 1/(\volcoh)^{1/3}$.

\begin{figure}[tbp]
\begin{center}
\includegraphics[width=.9\linewidth]{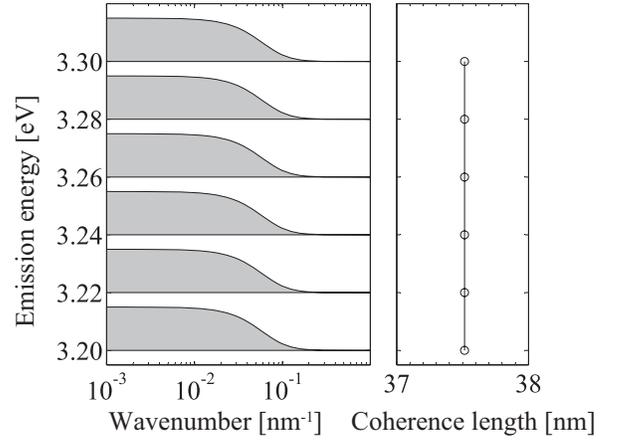}
\caption{(a) At each emission frequency,
the $k$-distribution of scattered exciton is plotted with gray color.
It is calculated from Eq.~\eqref{eq:S_calc} for ZnO
with fitting parameters $\lensct = 14.4\;\nm$ and $\lencoh^0 = 80\;\nm$.
(b) The coherence length $\lencoh$ of the scattered exciton
is calculated from the width at half maximum of the distribution.
At the present conditions, $\lencoh$ does not strongly depend
on the emission frequency.}
\label{fig:3}
\end{center}
\end{figure}
In Fig.~\ref{fig:3}(a), we calculated the $k$-distribution
of $S_{\xi,\vk}(\omega,0,0)$ in the following simple model,
and the estimated coherence length is plotted in Fig.~\ref{fig:3}(b).
The density of the bottleneck excitons is distributed as
a Gaussian function with a coherence length $\lencoh^{0}$as 
\begin{equation} \label{eq:Gauss} 
N_{1s,\xi,\vk} \propto \ee^{-(k\lencoh^{0}/2)^2}.
\end{equation}
The scattering coefficient does not depend on the exciton states
and is represented as
\begin{equation}
\sct_{\mu,\xi,\xi',\xi'',\vq} \propto \frac{1}{(q\lensct)^2 + 1}.
\end{equation}
Here, $\lensct$ is the screening length of the inelastic scattering,
which is expected to be in the order of the exciton Bohr radius,
which is $a_B^* = 1.8\;\nm$ for ZnO.\cite{Madelung2001}
The higher exciton states including the unbound ones are distributed continuously
above the band gap energy $E_g = \hbar\wx_{1s,k=0} + 60\;\meV$.\cite{Madelung2001}
The bound states below the band gap are not considered,
because the P emission is observed mainly for $\omega < \hbar\wx_{1s,k=0} - 60\;\meV$
in the experiments,
i.e., the contribution of the continuous band is dominant.
Then, we calculate the scattering coefficient as
\begin{align}
& S_{\xi,\vk}(\omega,0,0)
\nonumber \\ &
\propto \sum_{\vk',\vq} \int_{E_g/\hbar}^{\infty}\dd\omega'\
\frac{\ee^{-(|\vk+\vq|\lencoh^{0}/2)^2}\ee^{-(|\vk'-\vq|\lencoh^{0}/2)^2}}
     {[(q\lensct)^2 + 1]^2}
\nonumber \\ & \quad \times
\delta\left(\omega+\omega'+\hbar|\vk'|^2/2\mu
  - \wx_{1s,\vk+\vq}- \wx_{1s,\vk'-\vq} \right).
\label{eq:S_calc} 
\end{align}
Here, the reduced mass $\mu = (1/m_e + 1/m_h)^{-1}$
describes the dispersion of the continuous band,
while the frequency of the 1s exciton is
$\wx_{1s,k} = \wx_{1s,k=0} + \hbar k^2/2M$
for the total mass $M = m_e + m_h$
($m_e = 0.28m_0$ and $m_h = 0.59m_0$ in ZnO\cite{Madelung2001}).

The coherence length $\lencoh$ of the scattered excitons
plotted in Fig.~\ref{fig:3}(b)
is determined from the width at half maximum of the distribution
plotted in Fig.~\ref{fig:3}(a)
as the Gaussian distribution in Eq.~\eqref{eq:Gauss}.
The center-of-mass motion of the bottleneck exciton
has a finite coherence length $\lencoh^0$
and it is supposed to be longer than the screening length $\lensct \sim a_B^*$.
Under this condition, the coherence length $\lencoh$ of the scattered excitons
is basically determined by the screening length $\lensct$,
and $\lencoh$ is obtained generally shorter than
$\lencoh^0$ of the bottleneck excitons.
Due to the reabsorption problem,
it is hard to estimate $\lencoh^0$
from the free-exciton lifetime obtained in experiment.
Here, by supposing the values of $\lencoh^0$ and $\lensct$,
which are listed below,
we try to reproduce the coherence length of the scattered excitons
$\lencoh \sim 4\times10^1\;\nm$,
which was estimated in Sec.~\ref{sec:estimate_Vcoh} from the P-emission lifetime obtained experimentally.
In the calculation of Fig.~\ref{fig:3},
we supposed $\lensct = 8\times a_B^* = 14.4\;\nm$
as a main fitting parameter for obtaining $\lencoh \sim 40\;\nm$,
and $\lencoh^0 = 80\;\nm$ is chosen simply as twice this value
(e.g., we get $\lencoh \sim 20\;\nm$ for $\lensct = 4\times a_B^*$
and $\lencoh^0 = 80\;\nm$).
Under the present model and analysis,
we can only say that the coherence length of the bottleneck excitons $\lencoh^0$ 
should be longer than of the scattered excitons $\lencoh$
estimated from the P-emission lifetime.
Whereas such a long coherence length is expected
for the bottleneck exciton in our calculation,
its lifetime is elongated by the very small photonic fraction
and also by the dephasing process of the reabsorbed photons.
Note that, as seen in Fig.~\ref{fig:3}(b),
the coherence length $\lencoh$ of the scattered excitons
does not strongly depend on the emission frequency
at least under the present conditions.
Then, the P-emission lifetime at each emission frequency is basically
determined by the photonic fraction of the polariton state.

In general, the coherence volume
is determined through the last term in Eq.~\eqref{eq:ddt_ds_2}
or the last two terms in Eq.~\eqref{eq:master}
with self-consistently calculating $\orho(t)$ and $N_{\mu,\vk}(t)$.
Reflecting the coherence volume,
the density of scattered excitons
$\braket{\osigmad_{1s,\xi,\vk}\osigma_{1s,\xi,\vk}}$
initially spreads for $|\vk| \lesssim 1/(\volcoh)^{1/3}$.
After a long enough time compared to the P-emission lifetime,
the $\omega$-Fourier component of $\braket{\opd_{j,\vk,\eta}(t)\op_{j,\vk,\eta}(t')}$
should be distributed only around $k_L(\omega)$
reflecting the large coherence volume of the propagating polaritons.
Although we do not solve the master equation \eqref{eq:master} in this paper,
the exciton-to-polariton conversion time given by
such a calculation
should be equivalent to the one calculated in the previous section
if the coherence length just after the scattering
is shorter than the radiation wavelength.

At least theoretically, we can suppose any coherence volume
and emission frequency in the above calculation.
However, even by calculating the exciton-to-polariton conversion time
around the bottleneck region,
it is probably far from the spontaneous emission lifetime
observed in experiments.
The deviation basically originates from the two factors.
1) We must also consider the memory loss of the phase and propagation direction
of the reabsorbed photons
by considering the elastic exciton-exciton scattering,
interaction with phonons, etc.
2) The quasi-equilibrium at the bottleneck region
must be discussed under considering
the radiative recombination of exciton, reabsorption of the photon,
and the effect 1).
These problems are also remaining tasks in the future.

\subsection{Stimulated emission of polaritons or stimulated creation of excitons}
The P emission exhibits a threshold behavior with respect to the pumping power
and shows also an optical gain at that frequency.
\cite{Cingolani1982PRB,Tang1998APL,Ichida2000JL,Tanaka2002JAP,Nakayama2005APL,Hashimoto2010APL}
Then, the inelastic scattering has been considered
as a stimulated emission\cite{klingshirn05,Cingolani1982PRB,Tang1998APL,Ichida2000JL,Tanaka2002JAP,Nakayama2005APL,Hashimoto2010APL}
[or called the amplified spontaneous emission (ASE)]
and lasing is also reported.\cite{Tang1998APL}
In contrast, instead of the stimulated emission of photons or polaritons,
in this paper we interpret that the creation of excitons are stimulated
by the accumulated excitons with the P-emission energy
(stimulated scattering of excitons),
and then those excitons are emitted in the conversion time $\tconv$.

When we suppose that the polaritons are directly created
by the inelastic scattering, the deviation operator
in Eq.~\eqref{eq:ddt_ds_2} is approximated as
$\delta\osigma_{1s,\xi,\vk}(t')
\simeq \ee^{\ii\omega_{L,k}(t-t')}\sum_{\eta}X_{L,\vk,\eta,1s,\xi}\delta\op_{L,\vk,\eta}(t)$.
Then, the equation of motion of the number of polaritons
$\delta N_{L,\vk,\eta} = \braket{\delta\opd_{L,\vk,\eta}\delta\op_{L,\vk,\eta}}$ is obtained as
\begin{equation}
\ddt{}\delta N_{L,\vk,\eta}(t)
= - \gammabulk(\omega_{L,k}) \delta N_{L,\vk,\eta}(t)
+ G_{\vk,\eta} \delta N_{L,\vk,\eta}(t).
\label{eq:rate_polariton} 
\end{equation}
Here, the loss of the polaritons with the escape rate $\gammabulk(\omega)$,
Eq.~\eqref{eq:gamma_escape}, is introduced
and the gain $G_{\vk,\eta}$ is represented as
\begin{align}
G_{\vk,\eta}
& = \sum_{\mu\neq1s} \sum_{\xi,\xi',\xi''} \sum_{\vk',\vq}
    \frac{4\gammadepha X_{L,\vk,\eta,1s,\xi}{}^2\sct_{\mu,\xi,\xi',\xi'',\vq}{}^2}
         {(\delta\omega)^2+(2\gammadepha)^2}
\nonumber \\ & \quad \times
    \left[
          N_{1s,\xi',\vk'-\vq}N_{1s,\xi'',\vk+\vq}
  - (1+2N_{1s,\xi',\vk+\vq})N_{\mu,\vk'}
    \right].
\end{align}
Here, we simply supposed $\gammadepha = \gammadeph_{\mu,\vk}$.
The frequency difference
$\delta\omega = \omega_{L,k}+\wx_{\mu,\vk'}-\wx_{1s,\vk+\vq}-\wx_{1s,\vk'-\vq}$
gives a resonance at particular $\vk$
through the denominator $(\delta\omega)^2+(2\gammadepha)^2$.
Eq.~\eqref{eq:rate_polariton} corresponds to the rate equation
discussed in Sec.~22.1 of Ref.~\onlinecite{klingshirn05}.
When the gain becomes larger than the loss,
the stimulated emission of polaritons occurs,
and it determines the threshold of the P emission
in the conventional interpretation.

On the other hand, in our interpretation,
we suppose that the excitons are created by the inelastic scattering.
We define the density of scattered excitons $\delta N_{\vk,\xi}(\omega,t)$
converting to polaritons with a emission frequency $\omega$ as
\begin{equation}
\braket{\delta\osigmad_{\vk,\xi}(t')\delta\osigma_{\vk,\xi}(t)}
= \int\dd\omega\ \delta N_{\vk,\xi}(\omega,(t+t')/2) \ee^{-\ii\omega(t-t')},
\end{equation}
\begin{equation}
\delta N_{\vk,\xi}(\omega,t)
= \int\dd\tau\ \braket{\delta\osigmad_{\vk,\xi}(t-\tau/2)\delta\osigma_{\vk,\xi}(t+\tau/2)}
  \ee^{\ii\omega\tau}.
\end{equation}
The rate equation is derived from Eq.~\eqref{eq:ddt_ds_2} as
\begin{align}
\ddt{}\delta N_{\xi,\vk}(\omega,t)
& = - \varGamma(\omega) \delta N_{\xi,\vk}(\omega,t)
\nonumber \\ & \quad
+ 2\pi S_{\xi,\vk}(\omega,t,t) \delta N_{\xi,\vk}(\omega,t).
\label{eq:rate_exciton} 
\end{align}
For deriving this equation,
instead of considering explicitly the light-matter coupling,
the exciton-to-polariton conversion rate $\varGamma(\omega)$,
Eq.~\eqref{eq:Gamma(w)},
is introduced as the loss of the scattered excitons.
For deriving the second (gain) term,
we supposed that the dephasing rate 
is low enough than the oscillation frequency as $\gammadepha \ll \omega$,
and the density $N_{\xi,\vk}(\omega,t)$
is varying slowly with respect to $t$
compared to the dephasing time $1/\gammadepha$.
Also in our interpretation,
Eq.~\eqref{eq:rate_exciton} shows a threshold behavior
when the gain $2\pi S_{\xi,\vk}(\omega,t,t)$ exceeds the loss $\varGamma(\omega)$,
and a stimulated creation (scattering) of excitons occurs.

The stimulated emission of polaritons and the stimulated creation of excitons
are different processes with different thresholds.
In order to discuss theoretically which interpretation of the P emission is appropriate,
we should solve the master equation \eqref{eq:master}
without the assumptions of the direct creation of polaritons
or excitons by the inelastic scattering.
We do not perform such a calculation in this paper.
We instead justified our interpretation from the experimental results
\cite{Wakaiki2011pssc,Wakaiki2013EPJB,Wakaiki2014JL}
and from the discussion of the spontaneous emission from the bottleneck excitons,
in which the coherence volume plays an important role.
If the stimulated emission of photons or polaritons occurs
and the polaritons do not propagate diffusively,
we should observe the escape time $\tescape$ from the sample
as the P-emission lifetime.
However, the experimental data
\cite{Wakaiki2011pssc,Wakaiki2013EPJB,Wakaiki2014JL}
shows the lifetimes
one or two orders of magnitude slower than $\tescape$.
Further, since the excitons at the bottleneck are incoherent
(having a coherence length shorter than the radiation wavelength),
the scattered excitons are also supposed to have a poor coherence length.
They are the reasons why
we conclude that the stimulated creation of excitons occurs,
and those excitons are converted to polaritons
with the rate $\varGamma(\omega)$,
which is proportional to the group velocity $\vg(\omega)$
in the experiments and also in our calculation approximately.

In order to distinguish clearly the two stimulated processes experimentally,
we should perform a time-resolved measurement of the optical gain.
\cite{Cingolani1982PRB,Tang1998APL,Ichida2000JL,Tanaka2002JAP,Nakayama2005APL,Hashimoto2010APL}
We obtain the stimulated emission of polaritons
after the probe pulse arrives at the sample,
because the probe beam propagates as a polariton with a long enough spatial coherence.
\footnote{
When the lasing occurs without the probe pulse
and the radiation field gets a non-zero amplitude
with temporal and spatial coherences,
the inelastic scattering provides the stimulated emission of photons.
}
In our interpretation, the stimulated creation of excitons
occurs around the rise time of the P emission
(shortened inversely proportional to the square of the pump power
\cite{Ichida2005PRB}),
and the conversion from exciton to polariton occurs after that.
Then, the probe beam should get the optical gain
only in a time delay around the P-emission rise time
plus the onset time.
If the stimulated emission of photons or polariton occurs
and the created polaritons escape from the sample very quickly
in a time of $\tescape$,
the optical gain is obtained even in the decay period of the P emission,
because the P-emission lifetime corresponds to that of the bottleneck excitons.
Even if the polaritons propagate diffusively
and the stimulated emission of polaritons occurs only in the rise period,
the decay time of the optical-gain signal should reflect
the contribution of the relatively slow escape time of the diffusive polaritons.
In contrast, in our interpretation,
the decay time of the optical-gain signal should be shorter than
the P-emission lifetime, because spatial coherence is established
by the probe beam and the exciton-to-polariton conversion
is not restricted by the coherence volume.

\section{Summary} \label{sec:summary}
In the conventional interpretation of the P emission,
excitons at the bottleneck region are supposed to be scattered
directly to photon-like polariton states.
We instead propose another interpretation.
The excitons are scattered to bare exciton states first,
and then they are converted to polaritons
in a finite conversion time, which corresponds to the P-emission lifetime
observed in the recent experiments using the optical Kerr gating method.
\cite{Wakaiki2011pssc,Wakaiki2013EPJB,Wakaiki2014JL}
We justify our interpretation by supposing that
the scattered excitons should have a finite coherence volume
and they are converted to polaritons as the emission process from localized exciton.
Since the polariton states require a long enough spatial coherence
for their establishment,
they cannot be a direct destination of the inelastic scattering
because of the small coherence volume of the excitons.
In the calculation of the inelastic exciton-exciton scattering,
the coherence volume of the scattered excitons certainly appears
on the assumption that the bottleneck excitons originally have a finite coherence volume.
However, more detailed experimental and theoretical investigations
are required to finally conclude which interpretation is reasonable.
Especially, a time-resolved optical-gain measurement
would give us fruitful information
for distinguishing our interpretation,
the conventional one, and the one of the polariton diffusion.

\begin{acknowledgments}
M.B.~thanks B.~Deveaud for fruitful discussion.
This work was supported by JSPS KAKENHI (Grant No.~24560011, 26287087, and 24-632).
\end{acknowledgments}


\end{document}